\documentclass[11pt,twoside]{article}

\usepackage{asp2006}
\usepackage{epsf}
\usepackage{graphicx}
\usepackage{lscape}
\usepackage{amsbsy}
\usepackage{amssymb}

\begin{document}
\title{Spectropolarimetric diagnostics at the solar photosphere near the limb}

\author{L. Yelles Chaouche, S. K. Solanki}

\affil{Max Planck Institut f\"{u}r Sonnensystemforshung,
Max-Planck-strasse,2, Katlenburg-Lindau 37191, Germany}

\author{L. Rouppe van der Voort and M. van Noort}
 \affil{Institute of Theoretical Astrophysics, P.O. box
1029 Blindern N-0315, Oslo,  Norway}


\begin{abstract}
In the present work, we investigate the formation of Stokes profiles
and spectro-polarimetric diagnostics in an active region plage near
the limb. We use 3-D radiation-MHD simulations with unipolar fields
of an average strength of 400G, which is largely concentrated in
flux tubes in which the field reaches typical kilo-Gauss values. We
generate synthetic Stokes spectra by radiative transfer
calculations, then we degrade the simulated Stokes signal to account
for observational conditions. The synthetic data treated in this
manner are compared with and found to roughly reproduce
spectro-polarimetric high-resolution observations at $\mu$=0.39
obtained by the SOUP instrument with the Swedish 1-m Solar Telescope
at the beginning of 2006.

\end{abstract}


\section{Introduction}

The study of small scale magnetic flux concentrations near the solar
limb provides additional insight into the existing models of
magnetic flux concentrations \cite{Frutiger2003}. These flux
concentrations have been associated with solar faculae.
Understanding the physical processes behind solar faculae through
observations and simulations is an important topic
\cite{Lites2004,Keller2004,Carlsson2004,Steiner2005,Hirzberger2005,okunev2005,depontieu2006}.
One of the motivations for studying the facular phenomenon is the
influence of their brightness on the Sun's irradiance variation
\cite{fligge2001}.

In order to gain further insight into magnetic flux concentrations
in plage regions at intermediate $\mu$ values, we perform a study of
the spectropolarimetric signal in MHD simulations and observations
originating from plage regions at $\mu=0.39$. The simulations used
here are fully compressible 3-D radiation-MHD simulations
\cite{vog:etal2003,vog:etal2005}. The observations has been recorded
with the SOUP instrument \cite{Title1981} at the Swedish 1-m Solar
Telescope (SST) in 2006.

\section{Methods and simulations}

\begin{figure*} 
\includegraphics[width=0.7\textwidth]{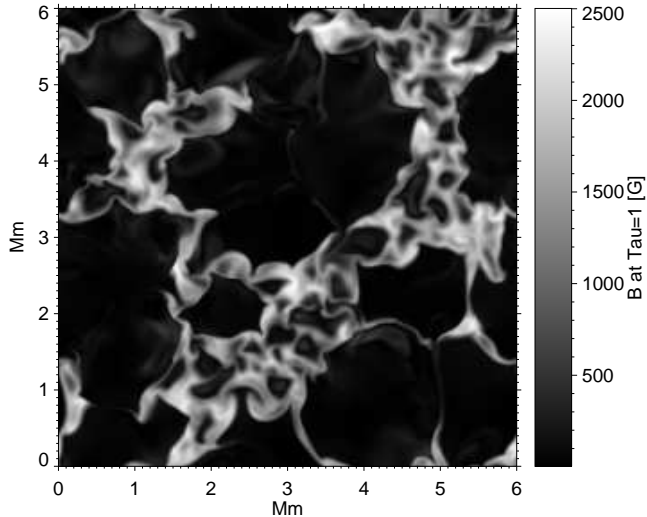}
\caption{Magnetic field strength map at an optical depth
$\tau_{5000}=1$. } \label{fig1}
\end{figure*}

\begin{figure*} 
\includegraphics[width=0.7\textwidth]{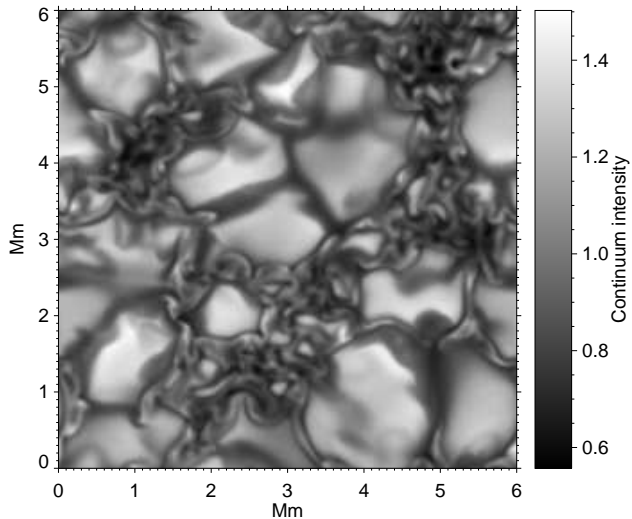}
\caption{Normalized continuum intensity map near $6302.5$ {\AA}
emerging from the simulation box at disc center. } \label{fig2}
\end{figure*}

The 3D radiation-MHD simulations used include, a solution of the
fully compressible MHD equations, including partial ionization in
the equation of state and a non-local non-grey radiative transfer
\cite{vog:etal2003,vog:etal2005}. The MHD simulation box is 6x6 Mm
in the horizontal plane and 1.4 Mm in the vertical direction, with a
resolution of 288x288x100 grid points. The mean magnetic field
strength is about 400 G. (See Figures~\ref{fig1} and ~\ref{fig2}).
In these Figures we can see the fully developed granulation pattern
which has interacted with the magnetic field. The flux gets
concentrated in intergranular lanes as a result of flux expulsion
\cite{schuessler1990}. The flux concentrations appear darker than
the average granular intensity. This is due to the partially
suppressed plasma motion through field lines. Nevertheless, a proper
analysis of energy exchange via convection and radiation is
necessary before drawing conclusions on the thermal, and radiative
properties of flux concentrations \cite{schuessler:etal2003}. This
is beyond the scope of the present paper.

\begin{figure*} 
\includegraphics[width=0.7\textwidth]{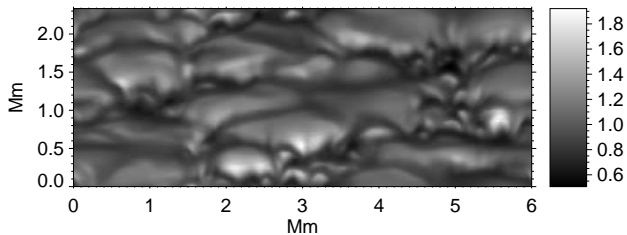}
\caption{Normalized continuum intensity map near $6302.5$ {\AA}
emerging from the simulation box at $\mu=0.39$. } \label{fig3}
\end{figure*}

We construct an inclined view of the simulation box by interpolation
of the different quantities along rays with the desired inclination
angle. This allows comparisons with observations away from disk
center. Fig.~\ref{fig3} shows an inclined view of Fig.~\ref{fig2} at
an angle of 67 degrees ($\mu=0.39$). One can notice the higher
values of intensity at some locations (e.g. at coordinates (2200km,
300km) and (3000km, 600km)). These are identified as faculae
\cite{Keller2004}.

In order to compare these 3D radiative-MHD simulations with
Spectropolarimetric observations, we have to calculate the Stokes
signal emerging from the inclined simulation box (e.g. the continuum
intensity map near 6302.5 {\AA} is shown in Figure~\ref{fig3}). This
is done using the STOPRO code in the SPINOR package \cite [Solanki
1987;] []{frutiger:etal2000}. In order to take into account effects
such as the finite aperture of the telescope, and seeing, we perform
then a spectral and spatial smearing of these synthetic polarimetric
data.

\section{Comparison with spectropolarimetric observations at Mu=0.39 }

\begin{figure}[h]
  \hfill
   \begin{center}
    \begin{minipage}[t]{.55\textwidth}

      \includegraphics[width=1.0\textwidth]{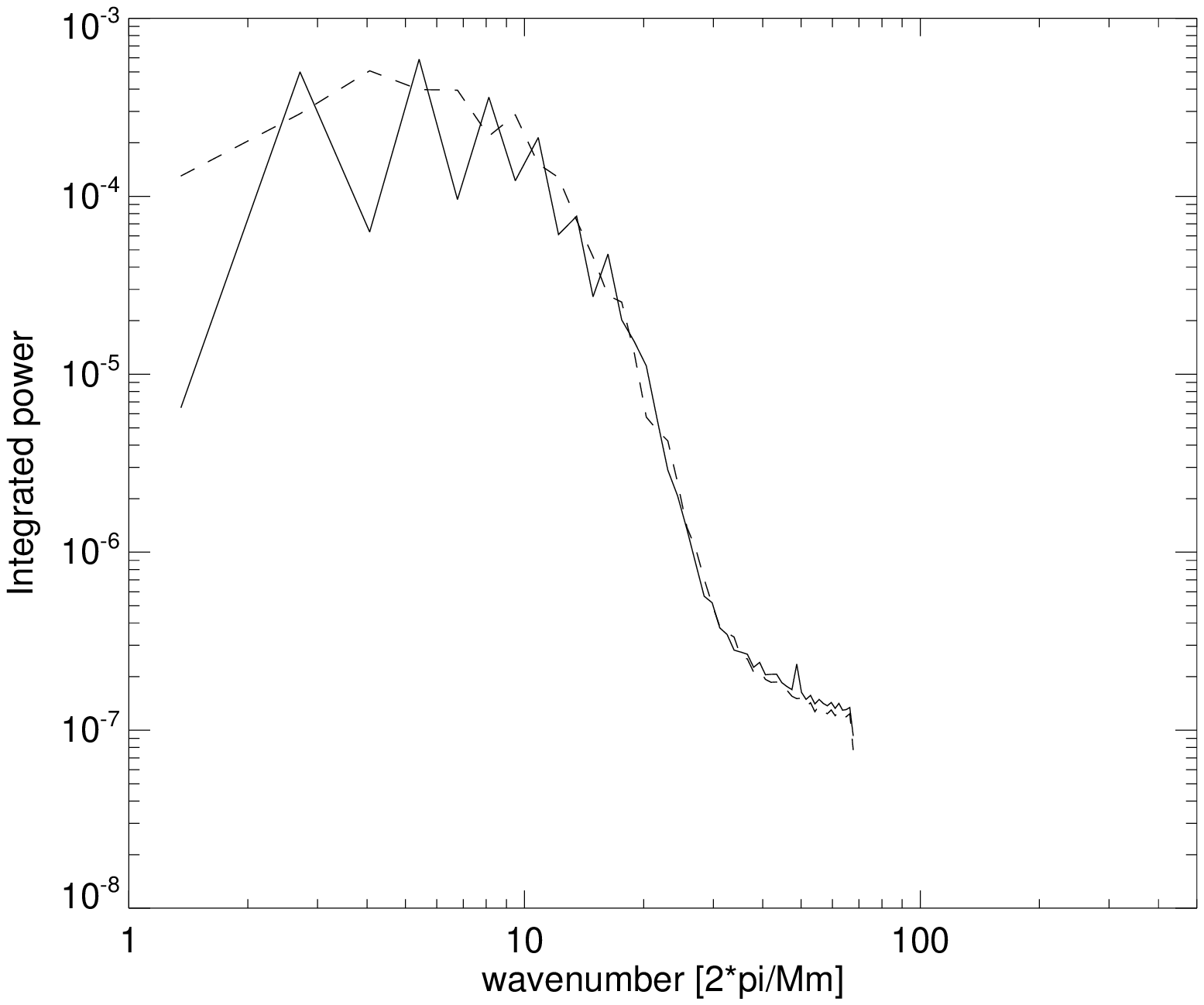}

    \end{minipage}
   \end{center}
  \hfill

  \hfill
  \begin{minipage}[t]{.45\textwidth}
    \begin{center}
      \includegraphics[width=1.0\textwidth]{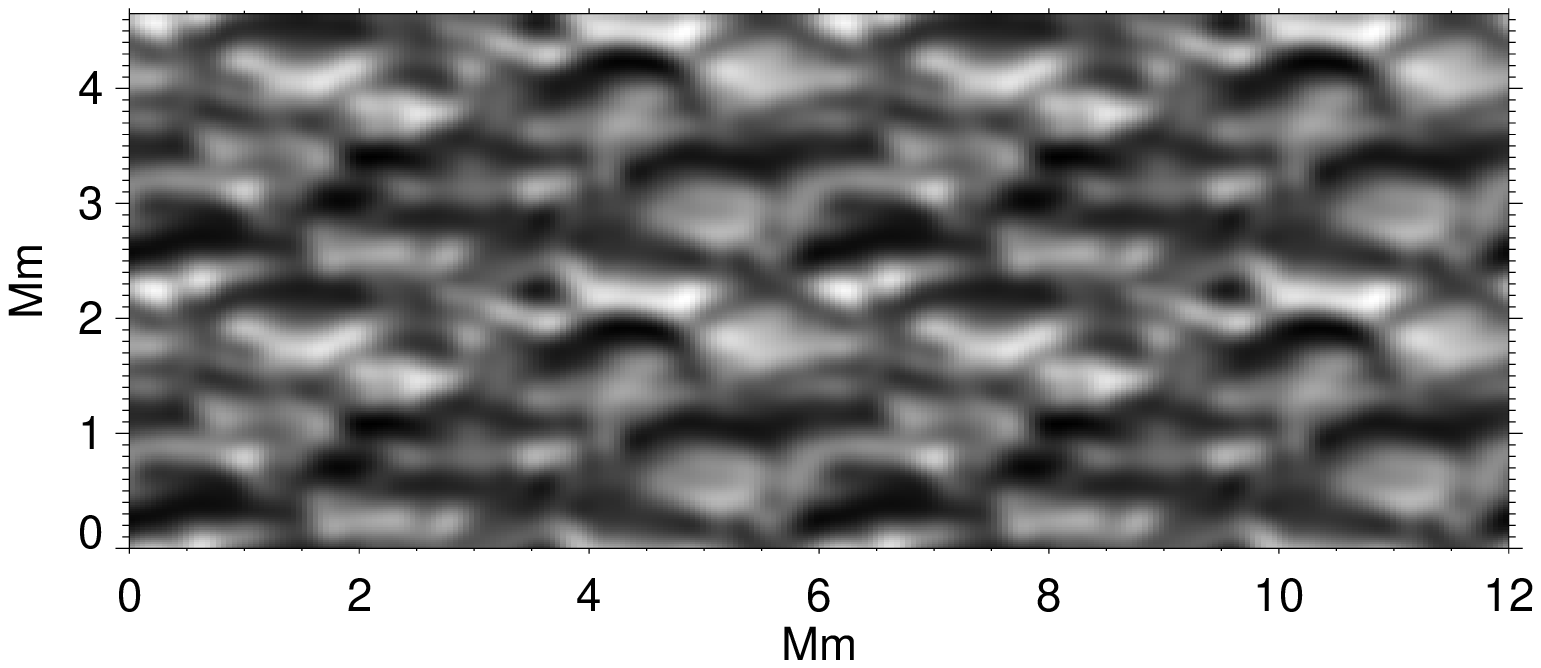}

    \end{center}
  \end{minipage}
  \hfill
  \begin{minipage}[t]{.45\textwidth}
    \begin{center}
      \includegraphics[width=1.0\textwidth]{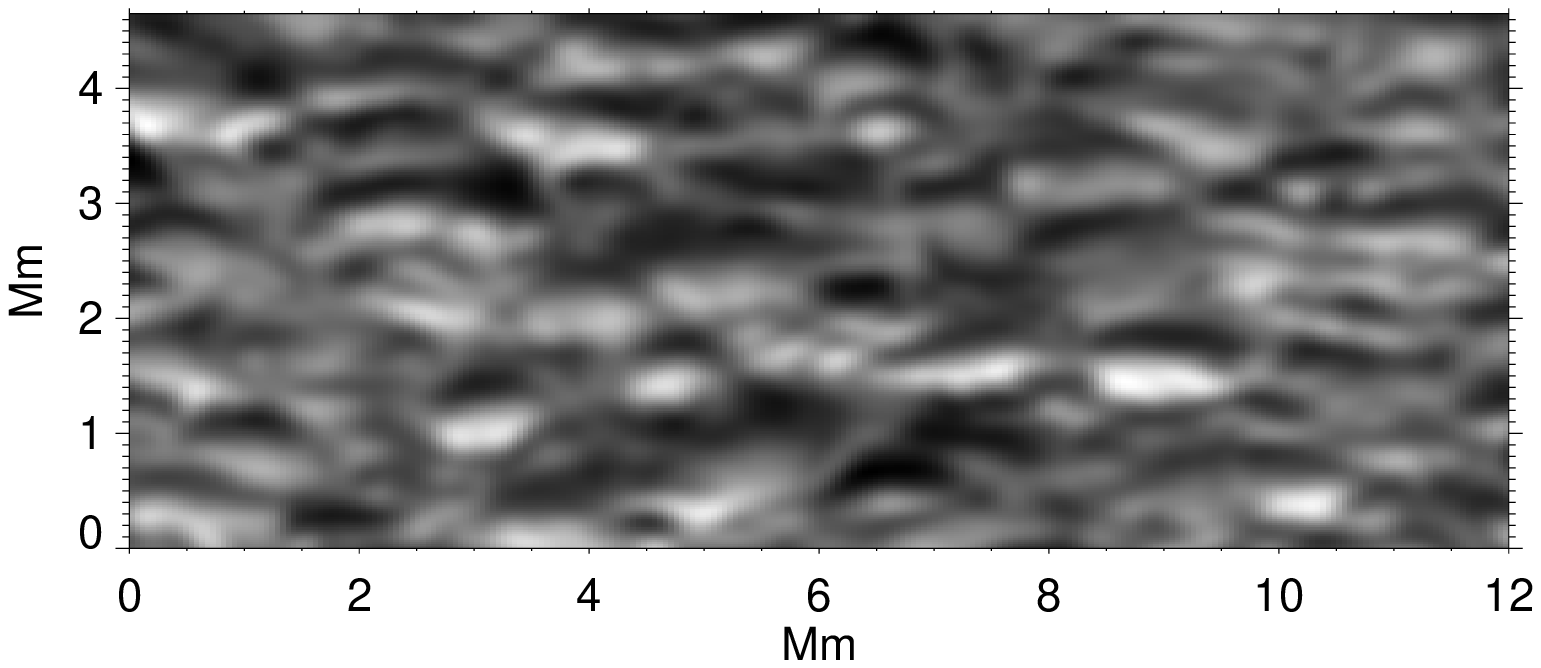}
    \end{center}
  \end{minipage}
  \hfill

       \caption{Two lower panels: Stokes-I maps at +50 m{\AA} from line
center Fe I 6302.5 {\AA}. Lower left: simulations, lower right :
observations. The upper panel shows the integrated power spectra for
the two lower images (full line : observations, dashed line:
simulations). }
     \label{fig4}
\end{figure}

  We compare the emerging Stokes signal from the simulation run with a
  set of high spatial resolution spectropolarimetric images obtained
  with the SOUP instrument at the Swedish 1-m Solar Telescope. The
  observational data consists of 9 images: the four Stokes polarization
  images at both + and -50 m{\AA} from the line center of the Fe I
  6302.5 {\AA}
  line, and a wide band image. The images are restored using the
  Multi-Object Multi-Frame Blind Deconvolution (MOMFBD) image
  restoration method \cite{vannoort:etal2005}. The SOUP images are based
  on 480 exposures at each line position which means that the effective
  exposure time amounts to 7.2 s. The combined use of the SST adaptive
  optics system and MOMFBD post-processing resulted in a spatial
  resolution in the Stokes observations that approaches the diffraction
  limit of the telescope (better than 0.2 arcsec).
In order to find the actual resolution of the polarimetric
observations, we carry out the following steps: 1/ We choose a
quiet-sun simulation run and convolve the synthetic Stokes profiles
in the spectral dimension with a Lyot type filter of FWHM=70 m{\AA}
(similar to SOUP). 2/ In order to account for the lower spatial
resolution in the
  observations, we apply a low-pass filter to the synthetic images
  which has the shape of a top-hat function and effectively removes
  power at the highest spatial frequencies - beyond the spatial
  resolution of the observations. In addition, we convolve the
  synthetic images with a Lorentzian profile which accounts for the far
  wings of the PSF that are not corrected for in the MOMFBD
  restoration. 3/ We pick a quiet region from the observed
  Stokes-$I$
images either at + 50 m{\AA} or -50 m{\AA} and compare it with the
synthetic ones. The matching of the two data sets is done by
comparing their power spectra (Figure ~\ref{fig4} upper panel) and
their standard deviations. This is done through an iterative process
where both the FWHM of the PSF and Lorenz profile are allowed to
change until the two data-sets fit with each other.


\begin{figure}[h]
  \hfill
  \begin{minipage}[t]{.45\textwidth}
    \begin{center}
      \includegraphics[width=1.0\textwidth]{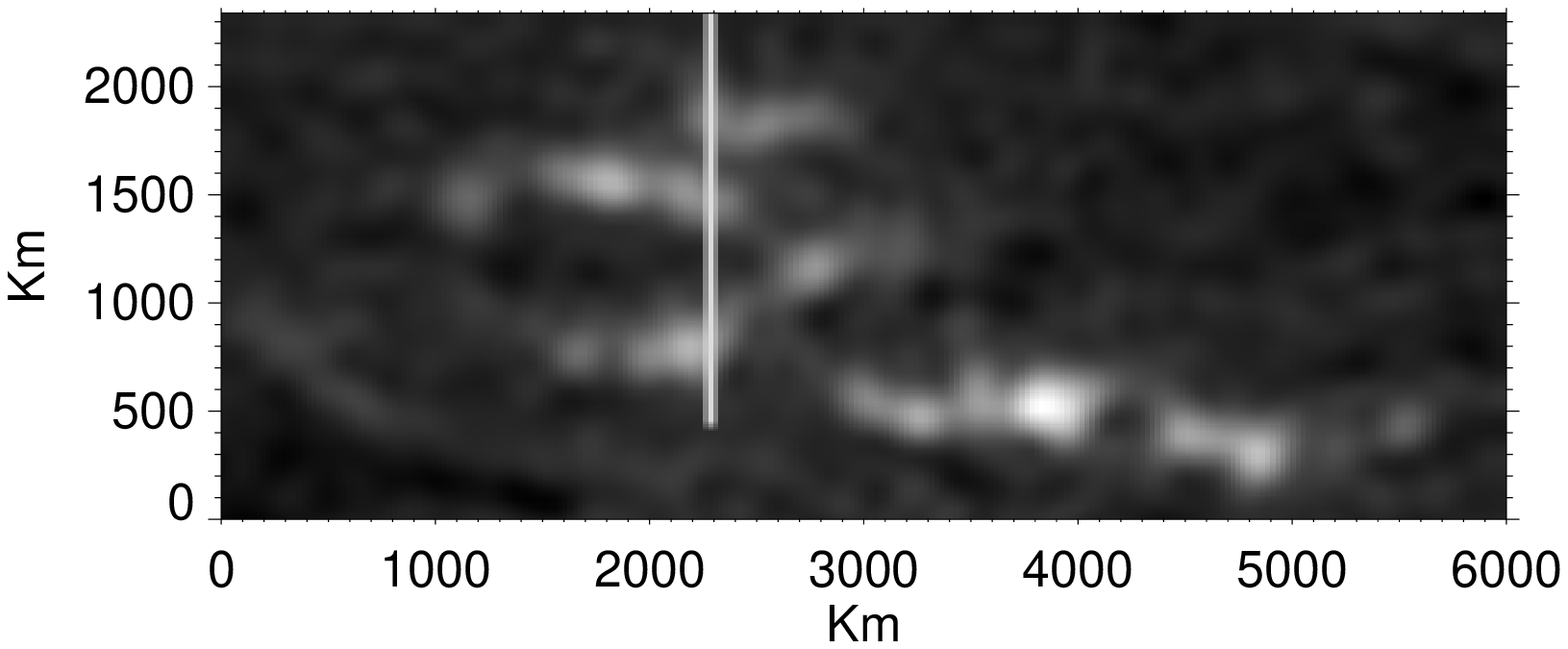}
    \end{center}
  \end{minipage}
  \hfill
  \begin{minipage}[t]{.45\textwidth}
    \begin{center}
      \includegraphics[width=0.9\textwidth]{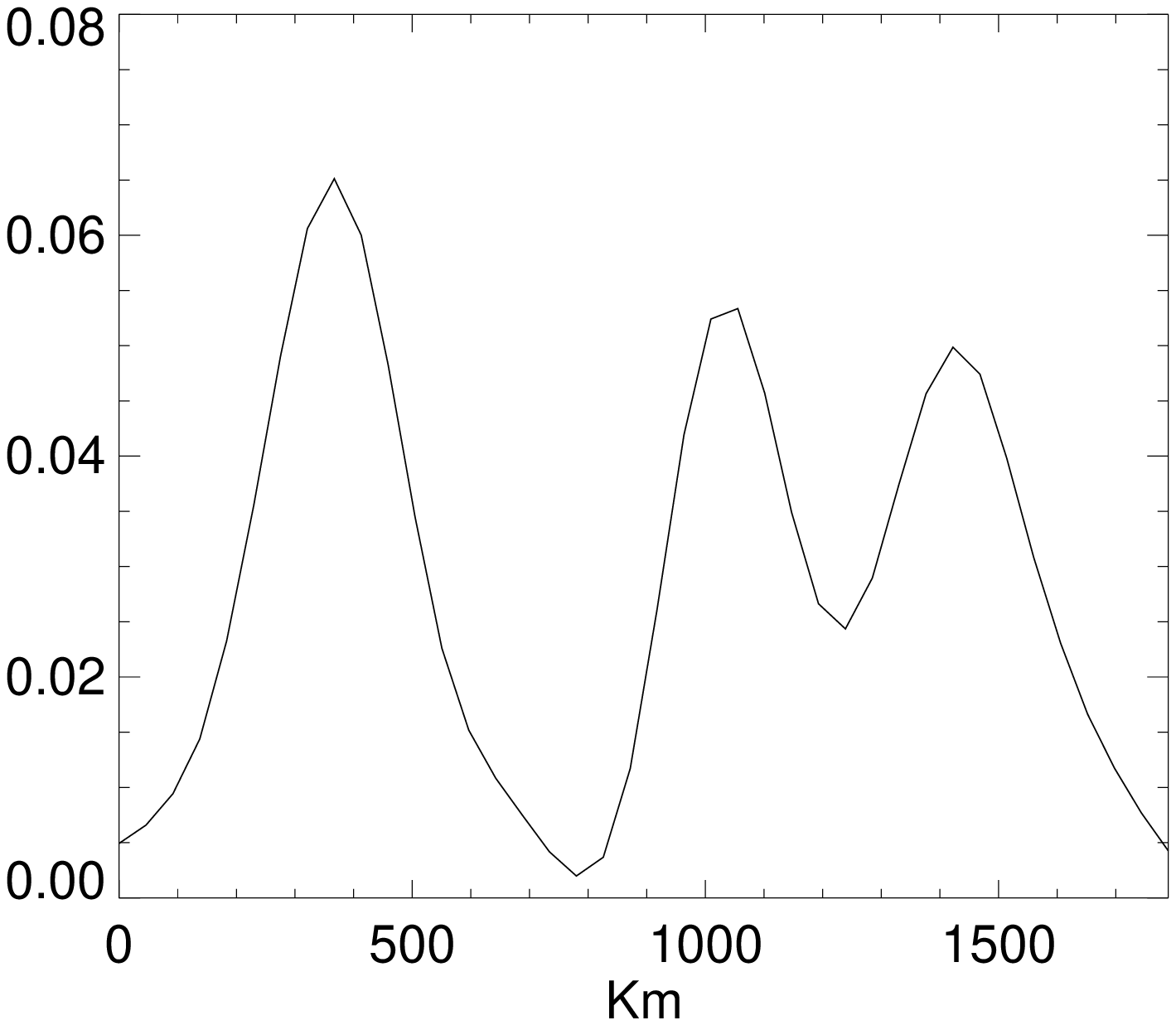}
    \end{center}
  \end{minipage}
  \hfill

  \hfill
  \begin{minipage}[t]{.45\textwidth}
    \begin{center}
      \includegraphics[width=1.0\textwidth]{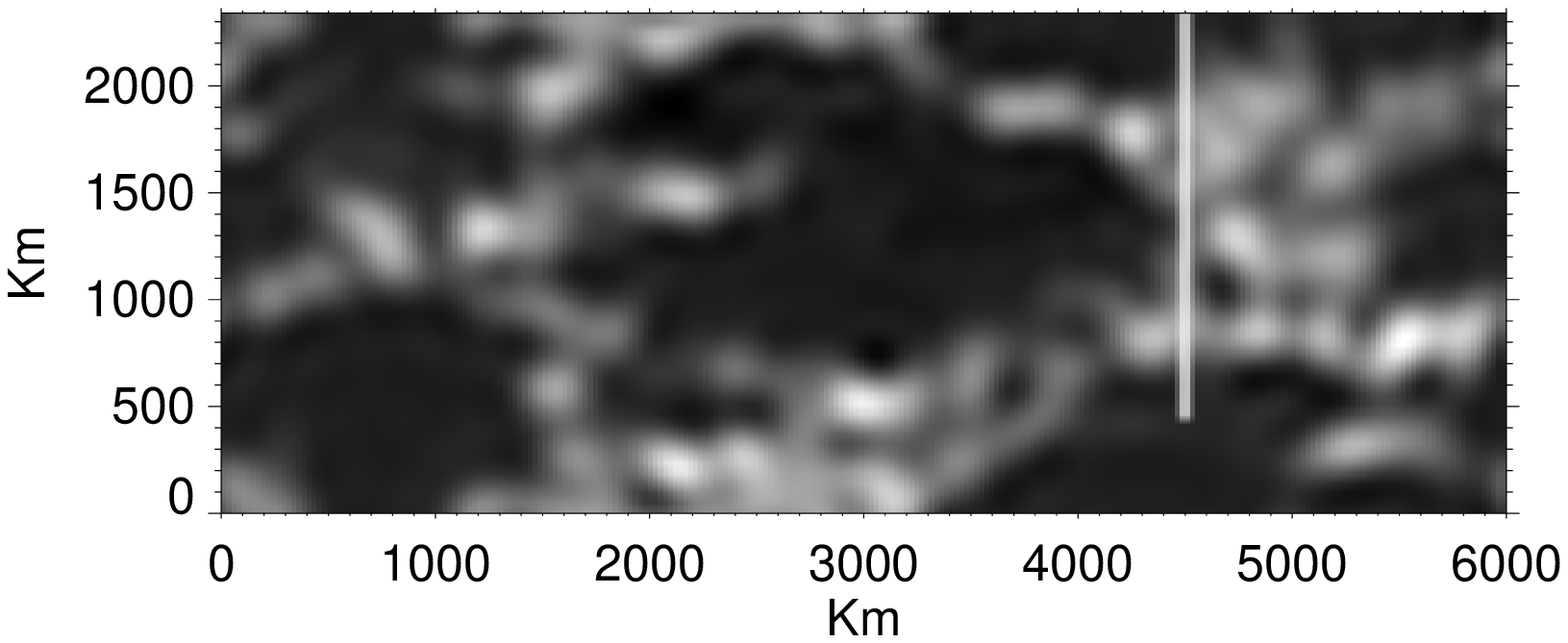}

    \end{center}
  \end{minipage}
  \hfill
  \begin{minipage}[t]{.45\textwidth}
    \begin{center}
      \includegraphics[width=0.9\textwidth]{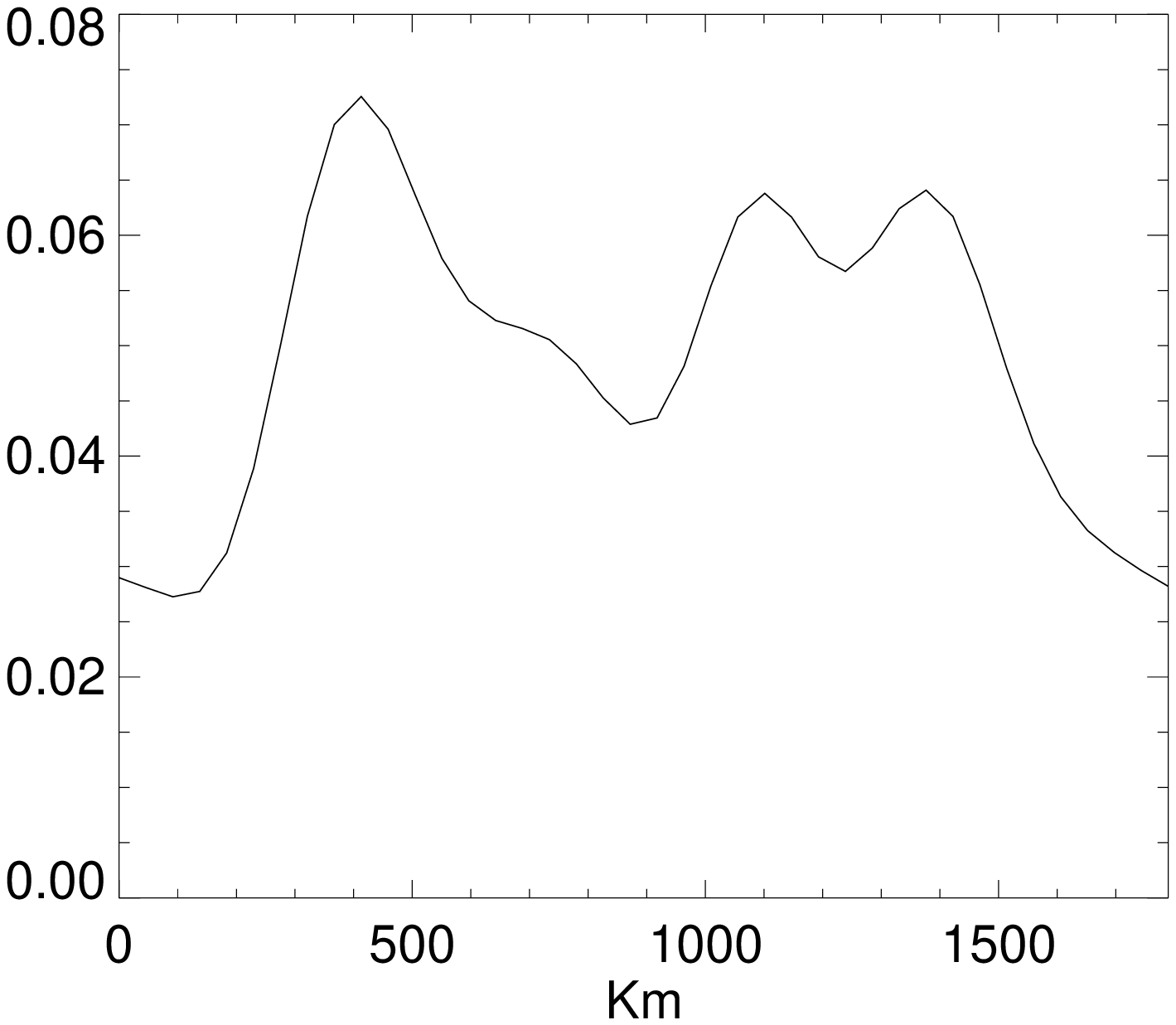}
    \end{center}
  \end{minipage}
  \hfill

       \caption{Upper left panel: observed Stokes-V at -50 m{\AA} from line
center Fe I 6302.5 {\AA}. Lower left panel: Stokes-V from
simulations degraded to the resolution of observations. A measure of
the Stokes-V signal at the white slits-positions is shown on the
right panels. }
     \label{fig5}
\end{figure}


The so obtained degrading parameters are used when comparing more
active regions, like the ones in Figure~\ref{fig5}.  The two lower
panels of Figure~\ref{fig4} indicate the observed (right one) and
degraded-simulated (left one) Stokes-I maps at +50 m{\AA} from line
center (Fe I 6302.5 {\AA}). The simulated map corresponds to a quiet
region where the mean field strength is about 20G. One can see the
similarity of the two images. The original observed image was
inclined by an angle of about $20$ degrees, which we have corrected
here.

In the two left images in Figure~\ref{fig5}, we see an observations
and a degraded simulation of Stokes-V at -50 m{\AA}. A direct
quantitative comparison between Stokes-V signals in these two images
can be done by plotting the signal along slits (e.g. the white slits
in the left panels). The resulting signal is shown in the two right
panels. We observe that the two curves are reasonably similar. This
indicates that the observed features here are well reproduced by an
MHD simulation run with an average field strength of 400G. At this
stage we can address the question: does the MHD simulation-run
reproduce the actual photospheric situation at the observed region
on figure~\ref{fig5} ?. The results above are in favor of a positive
answer. But the fact that we have only two spectral positions at +50
m{\AA} and -50 m{\AA} makes it necessary to proceed to further
comparisons with a larger spectral sampling. Then, in case
simulations and observations match, we can proceed further by
studying these regions in detail directly from the MHD simulations.

\section{Conclusions}

We presented here a comparative study about small-scale magnetic
flux concentrations near the solar limb. This is part of a more
general work aiming at investigating their physical and
spectropolrimetric properties through Stokes-diagnostics and direct
MHD simulations. We focussed here on identifying similarities
between observed and simulated Stokes signal at + and -50 m{\AA}
from line center of Fe I 6302.5 {\AA}. The comparison has been made
possible, by introducing suitable instrumental and seeing
degradation to the simulated Stokes signal. We thus identified
similarities between the observed and simulated maps of Stokes-V at
two wavelength positions + and -50 m{\AA} from line center of Fe I
6302.5 {\AA}. Although we see similarities, we mention the necessity
of further investigation before concluding that the simulations are
indeed reproducing the observed magnetic features. Such
investigations should include more spectral positions, allowing a
more consistent study of the variation of physical quantities (e.g.
temperature, magnetic field vector, line-of-sight velocity) through
the inclined magnetic flux concentrations.



\end{document}